\documentstyle[12pt]{article}

\def\scrip{{\cal I^+}}

\def\w{{\omega}}

\def\singlespace {\smallskipamount=3.75pt plus1pt minus1pt
                  \medskipamount=7.5pt plus2pt minus2pt
                  \bigskipamount=15pt plus4pt minus4pt
                  \normalbaselineskip=15pt plus0pt minus0pt
                  \normallineskip=1pt
                  \normallineskiplimit=0pt
                  \jot=3.75pt
                  {\def\smallskip {\vskip\smallskipamount}}
                  {\def\medskip   {\vskip\medskipamount}}
                  {\def\bigskip   {\vskip\bigskipamount}}
                  {\setbox\strutbox=\hbox{\vrule 
                    height10.5pt depth4.5pt width 0pt}}
                  \parskip 7.5pt
                  \normalbaselines}
\def\middlespace {\smallskipamount=5.625pt plus1.5pt minus1.5pt
                  \medskipamount=11.25pt plus3pt minus3pt
                  \bigskipamount=22.5pt plus6pt minus6pt
                  \normalbaselineskip=22.5pt plus0pt minus0pt
                  \normallineskip=1pt
                  \normallineskiplimit=0pt
                  \jot=5.625pt
                  {\def\smallskip {\vskip\smallskipamount}}
                  {\def\medskip   {\vskip\medskipamount}}
                  {\def\bigskip   {\vskip\bigskipamount}}
                  {\setbox\strutbox=\hbox{\vrule 
                    height15.75pt depth6.75pt width 0pt}}
                  \parskip 11.25pt
                  \normalbaselines}
\def\doublespace {\smallskipamount=7.5pt plus2pt minus2pt
                  \medskipamount=15pt plus4pt minus4pt
                  \bigskipamount=30pt plus8pt minus8pt
                  \normalbaselineskip=30pt plus0pt minus0pt
                  \normallineskip=2pt
                  \normallineskiplimit=0pt
                  \jot=7.5pt
                  {\def\smallskip {\vskip\smallskipamount}}
                  {\def\medskip   {\vskip\medskipamount}}
                  {\def\bigskip   {\vskip\bigskipamount}}
                  {\setbox\strutbox=\hbox{\vrule 
                    height21.0pt depth9.0pt width 0pt}}
                  \parskip 15.0pt
                  \normalbaselines}

\begin{document}

\begin{center}
{\bf {\Large Radiation Flux and Spectrum in the}}

{\bf {\Large Vaidya Collapse Model}}

\bigskip\ 

{\bf T. P. Singh\footnote{{ e-mail address:
tpsingh@tifr.res.in}}}

{\it Tata Institute of Fundamental Research,}

{\it Homi Bhabha Road, Mumbai 400 005, India.}

\smallskip\ 

{\bf Cenalo Vaz\footnote{{ e-mail address: cvaz@ualg.pt}}}

{\it Unidade de Ciencias Exactas e Humanas}

{\it Universidade do Algarve, Faro, Portugal}

\bigskip

\end{center}

\begin{abstract}
We consider the quantization of a massless scalar field, using the geometric
optics approximation, in the background spacetime of a collapsing spherical 
self-similar Vaidya star, which forms a black hole or a naked singularity. We
show that the outgoing radiation flux of the quantized scalar field diverges
on the Cauchy horizon. The spectrum of the produced scalar partcles is 
non-thermal when the background develops a naked singularity. These results 
are analogous to those obtained for the scalar quantization on a self-similar
dust cloud.  
\end{abstract}

\middlespace

\section{Introduction}

If the classical gravitational collapse of a star ends in a naked
singularity, then it is of interest to know how the quantum evaporation of
such a star (via particle creation) compares with the quantum evaporation
(via Hawking radiation) of another star that ends as a black hole. A few
studies have been carried out in recent years, with an attempt to answer
this question. These studies can be traced back to the pioneering works of
Ford and Parker \cite{fpa}, and of Hiscock et al. \cite{his}. 

Ford and Parker studied the outgoing quantum flux of a massless scalar
field, using the geometric optics approximation, in the background geometry
of a classical, spherical dust star forming a shell-crossing naked
singularity. They found that this flux does not diverge in the approach to
the Cauchy horizon. Their method was used by Barve et al. 
\cite{ba1} to calculate the
quantum flux in the geometry of a spherical self-similar dust cloud which
develops a shell-focusing naked singularity - in this model the flux
diverges on the Cauchy horizon, in a positive sense.

Hiscock et al. studied the outgoing flux in the self-similar Vaidya
model by obtaining the stress-energy tensor in the effective 2-d Vaidya
background from the trace anomaly and the conservation equations.  This
calculation is in the same spirit as the geometric optics approximation in the
sense that only the lowest angular momentum modes are accounted for. 
However, it has the advantage of yielding the components of the stress tensor
everywhere in the spacetime and not just asymptotically. Once again they
obtained a divergent flux on the Cauchy horizon. Barve et al.
\cite{ba2} repeated their analysis for the self-similar dust collapse model 
and obtained similar results. 

A new facet was added to these studies by Vaz and Witten
\cite{vaw}, who demonstrated how
one could calculate the spectrum of the outgoing radiation, in the approach
to the Cauchy horizon, for the self-similar dust model. The spectrum was 
shown to be non-thermal, and this is a characteristic feature distinguishing
the evaporation of naked singularities from thermal black hole evaporation.
A detailed justification on the use of Bogoliubov transformations in the
presence of a Cauchy horizon has been given by us recently \cite{vas}.

In the present paper, we apply the geometric optics approximation to the
quantization of a massless scalar field in the background of a 
spherical collapsing self-similar Vaidya star. Such a star results in
black hole formation for part of the initial data, and a naked singularity  
for the remaining initial data. In this approximation, we calculate the
flux of the scalar field emitted to infinity, and the spectrum of the emitted
radiation. The essential quantity needed for these calculations is the map
from an ingoing null ray to an outgoing null ray which passes through the
center of the collapsing cloud. This map is found in Section 2. In Section 3 we
show that the flux diverges on the Cauchy horizon, and that the spectrum of the
radiation is non-thermal.

These results are identical to those that have been found earlier for the
model of self-similar dust collapse \cite{ba1}, \cite{vaw}.
They provide additional evidence for the
conjecture that the divergence of the outgoing flux on the Cauchy horizon 
and a non-thermal radiation spectrum are generic features of naked 
singularities.

\section{The Classical Solution}

In this Section, we find the mapping from an ingoing null ray to an outgoing
null ray, for those initial conditions which result in a naked singularity
or black hole, in the self-similar Vaidya model. The metric is given by 
\begin{equation}
\label{met}ds^2=\left( 1-\frac{2m(v)}R\right) dv^2-2dvdR-R^2d\Omega ^2
\end{equation}
where $v$ is the advanced time coordinate $(-\infty <v<\infty )$ and $R$ is
the area radius $(0\leq R<\infty )$. The mass function is zero for $v<0$, it
is $m(v)=\mu v$ for $0<v<v_0$, and equal to a constant $M=\mu v_0$ for $%
v\geq v_0$. This represents a collapsing ball of null dust bounded in the
region $0<v\leq v_0.$ The linearity of the mass function ensures that this
Vaidya spacetime is self-similar, i.e. it possesses a homothetic Killing
vector field. Clearly, spacetime is Minkowski for $v<0$, and here the metric
may be written as 
\begin{equation}
\label{fla}ds^2=dUdv-R^2d\Omega ^2
\end{equation}
where $U=v-2R$ is the retarded time coordinate. Outside the Vaidya region,
the geometry is Schwarzschild, and in terms of the Eddington-Finkelstein
null coordinates, is given by 
\begin{equation}
\label{Sch}ds^2=\left( 1-\frac{2M}R\right) d\hat Udv-R^2d\Omega ^2,
\end{equation}
with $\hat U=v-2R^{*}$, and $R^{*}=R+2M\ln \left( R/2M-1\right) $.

We now obtain double null coordinates in the Vaidya region, by first
defining $x=v/R$ and $z=\ln R,$ so that the metric (\ref{met}) becomes 
\begin{equation}
\label{met2}ds^2=\exp (2z)\,\left[ A(x)dx^2+B(x)dz^2+2C(x)dxdz\right] . 
\end{equation}
Here, 
\begin{equation}
\label{abc}A(x)=1-2\mu x,\quad B(x)=\left( 1-2\mu x\right) x^2-2x,\quad
C(x)=\left( 1-2\mu x\right) x-1. 
\end{equation}
Next, we define 
\begin{equation}
\label{dta}d\tau =dz+\frac{C(x)}{B(x)}\,dx, 
\end{equation}
and 
\begin{equation}
\label{dch}d\chi =-\left[ \frac{C^2(x)-A(x)B(x)}{B^2(x)}\right] ^{1/2}dx, 
\end{equation}
so that the metric can be written as 
\begin{equation}
\label{met3}ds^2=\exp (2z)\,B(x)\,\left[ d\tau ^2-d\chi ^2\,\right] . 
\end{equation}
We define the double-null coordinates 
\begin{equation}
\label{nul}u=\exp (\tau -\chi ),\qquad v=\exp (\tau +\chi ), 
\end{equation}
so that 
\begin{equation}
\label{met4}ds^2=\frac{\exp (2z)\,B(x)}{uv}dudv. 
\end{equation}

By using the identity $C^2(x)-A(x)B(x)=1$, the null coordinates may be
written as 
\begin{equation}
\label{null}u=r\exp I_{+}\,,\qquad v=r\exp I_{-} 
\end{equation}
where 
\begin{equation}
\label{int}I_{\pm }=\int \frac{C(x)\pm 1}{B(x)}dx. 
\end{equation}
It is easily shown that $I_{-}=\ln x$, so that $v$ is trivially the same as
the advanced time coordinate introduced in the beginning. We have chosen to
write the null coordinates in this manner essentially to show that they can
be constructed in a symmetric fashion. The integral $I_{+}$ can also be
easily carried out, and we get 
\begin{equation}
\label{nul2}u=R\,\left( x-\alpha _{+}\right) ^{A_{+}}\;(x-\alpha
_{-})^{A_{-}}\quad , 
\end{equation}
where 
\begin{equation}
\label{alp}\alpha _{\pm }=\frac{1\pm \left( 1-16\mu \right) ^{1/2}}{4\mu } 
\end{equation}
are the roots of the quadratic equation 
\begin{equation}
\label{qua}x^2-\frac x{2\mu }+\frac 1\mu =0, 
\end{equation}
and the coefficients $A_{+}$ and $A_{-}$ are given by 
\begin{equation}
\label{coe}A_{+}=-\frac{1-2\mu \alpha _{+}}{\alpha _{+}-\alpha _{-}},\qquad
A_{-}=-\frac{1-2\mu \alpha _{-}}{\alpha _{-}-\alpha _{+}}. 
\end{equation}
We note that $A_{+}+A_{-}=1$. Also, the ingoing null coordinate used by
Hiscock et al. \cite{his} is obtained by taking the logarithm of the 
coordinate $u$ defined in (\ref{nul2}).

The flat spacetime limit of the Vaidya metric is obtained by setting $\mu =0$%
. In this limit, $v=t+R$, and it is easily shown that in this limit the
coordinate $u$ reduces to $t-R$.

The roots $\alpha _{\pm }$ are real for $\mu \leq 1/16$ and complex for $\mu
>1/16$. We note that when the roots are complex, $u$ continues to be real,
and can be written as 
\begin{equation}
\label{uim}u=R\,|x-\alpha _{+}|^2\;\exp (-2\xi ImA_{+}) 
\end{equation}
where $\xi $ is the phase of $\alpha _{+}$.

As is known, the collapse of the cloud results in the formation of a
curvature singularity, when the area radius of a shell shrinks to zero. The
Kretschmann scalar is given by 
\begin{equation}
\label{kre}R^{\alpha \beta \gamma \delta }R_{\alpha \beta \gamma \delta }=%
\frac{48m^2(v)}{R^6}=48\mu ^2x^6v^{-4}. 
\end{equation}
The central singularity forms when the radius of the innermost shell, given
by $v=0$, becomes zero. The curvature diverges when the center $R=0$ is
approached along the line $v=0$, with $x=v/R$ a finite non-zero constant. It
is known \cite{his}
that this singularity is naked for $\mu \leq 1/16$ and covered for $%
\mu >1/16$. The singularity corresponding to all shells with $v>0$ is
covered.

The nature of the central singularity, i.e. as to whether it is naked or
not, can also be deduced from the construction of the double null
coordinates given above. We note from (\ref{nul2}) that for $v<0$ the center
(i.e. $R=0$) is given by the line $u=v$. Let us assume that the singular
point $v=0,R=0$ is approached along the line $x=x_0$, with $x_0$ a finite,
non-zero constant. Then, from (\ref{uim}) we see that the ratio $u/v$ is
given by 
\begin{equation}
\label{rat}\frac uv=d=\frac{|x_0-\alpha _{+}|^2\;\exp (-2\xi ImA_{+})}{x_0} 
\end{equation}
with $d$ in general not equal to one. The singular point $v=0,R=0$ is thus
the intersection of two lines, $u=v$ and $u=dv$. Equivalently, it is given
by $u=v=0$.

If the central singularity is naked, there will be at least one outgoing
null ray from the singularity. This means that such a ray, for which $u=0$,
has $R\neq 0$. From (\ref{nul2}) we see that this is possible if and only if 
$x=\alpha _{+}$ or $x=\alpha _{-}$, with $\alpha _{\pm }$ real. The
existence of a real root is necessary for the central singularity to be
naked, implying that $\mu \leq 1/16$. The Cauchy horizon, which is the first
null ray to leave the singularity, is given by $x=\alpha _{-}$, where $%
\alpha _{-}$ is the smaller of the two roots. The Eddington-Finkelstein null
coordinate $\hat U$ corresponding to this ray is found by noting that when
the ray meets the boundary $v_0$ of the star, $R=v_0/\alpha _{-}$. Using
this value of $R$ in the definition of $\hat U$ we get 
\begin{equation}
\label{bou}\hat U=v_0-\frac{2v_0}{\alpha _{-}}-4M\ln |\frac{v_0}{2M\alpha
_{-}}-1|. 
\end{equation}
This value is not infinite, which shows that the singularity is globally
naked.

We can now work out the map 
${\cal F}(v)$ between an ingoing null ray and the outgoing
null ray to which it corresponds, i.e. we can find the function 
$\hat U={\cal F}(v)$%
, for both the black hole case and the naked singularity case. In the naked
case, we do this for outgoing rays with a value of $u$ close to zero.
Consider a null ray, incoming from $\cal{I^{-}}$, given by $v=$ constant,
with $v<0$. After passing through the center, this becomes an outgoing ray $u$,
with $u=v$. For $u$ nearly zero, $x$ is close to $\alpha _{-}$, so we write
$x=\alpha _{-}+\hat x$. From (\ref{nul2}) we can write 
\begin{equation}
\label{uou}u=BR\hat x^{A_{-}},\qquad B=(\alpha _{-}-\alpha _{+})^{A_{+}}. 
\end{equation}
By noting that 
\begin{equation}
\label{aar}R=\frac{v_0}{\alpha _{-}+\hat x^{}}\simeq \frac{v_0}{\alpha _{-}}%
\left( 1-\frac{\hat x}{\alpha _{-}}\right) 
\end{equation}
and by using this relation in (\ref{uou}) we get 
\begin{equation}
\label{exx}\hat x=\left( \frac{\alpha _{-}}{Bv_0}\right)
^{1/A_{-}}u^{1/A_{-}}. 
\end{equation}
Using these expressions for $R$ and $\hat x$ in the definition of $\hat U$
we can conclude that 
\begin{equation}
\label{map}\hat U=\hat U^0-Q\,v^{1/A_{-}}={\cal F}(v) 
\end{equation}
where $\hat U^0$ is the Cauchy horizon, and $Q$ is an (irrelevant) constant.
This is the desired map in the naked singularity case. The inverse function, 
$v={\cal G}(\hat U)$ is given by 
\begin{equation}
\label{inv}v={\cal G}(\hat U)=\left( \frac{\hat U^0-\hat U}Q\right) ^{A_{-}}. 
\end{equation}

In order to find the function 
${\cal F}(v)$ for the case in which the collapse ends
in a black hole (i.e. $\mu >1/16$), we examine rays which arrive at 
${\cal I^{+}}$
just before the event horizon. The event horizon is the outgoing ray which
reaches the boundary $v_0$ when the boundary has an area radius $R=2M$,
which implies that $x=v_0/2M$ for such a ray. By writing $x=\hat x+v_0/2M$,
we easily note that to leading order $\hat U=-4M\ln \hat x$ for such a ray.
Further, since $\hat x$ is small, $u$ is of the form $u=a+b\hat x$, and
using $u=v$ for a ray that goes through the center, we can conclude that 
\begin{equation}
\label{ma2}\hat U=-4M\ln \left( \frac{v-a}b\right) ={\cal F}(v) 
\end{equation}
which is the map in the black hole case. The inverse function, 
$v={\cal G}(\hat U)$
is now given by 
\begin{equation}
\label{in2}v={\cal G}(\hat U)=a+be^{-\hat U/4M}. 
\end{equation}

\section{The Quantum Flux}

We consider next the quantization of a massless scalar field $\phi (x)$ in
the background geometry of the collapsing Vaidya cloud. We will calculate
the flux of the energy radiated to 
${\cal I^{+}}$, and the spectrum of the emitted
radiation, in the geometric optics approximation. The flux radiated to
infinity is given by the off-diagonal component of the stress-energy tensor of
the massless scalar field, as \cite{fpa}
\begin{equation}
\label{flu}
P~~ =~~ \int  {}_{M}\langle 0 | T^R_T | 0 \rangle_{M}
 R^2 \sin\theta d\theta d\phi~~ 
=~~ {1 \over
{24\pi}} \left[{{{\cal F}{'''}} \over {({\cal F}')^3}}~ -~ 
{3 \over 2} \left({{{\cal F}
{''}} \over {{\cal F}^{'2}}} \right)^2\right]
\end{equation}
where the function ${\cal F}(v)$ has been calculated above. 
$|0\rangle_{M}$ is the Minkowski vacuum.
For the naked
singularity case we substitute the expression (\ref{map}) for 
${\cal F}(v)$ and
find the radiated power to be 
\begin{equation}
\label{npo}P(v)=\frac 1{48\pi Q^2}\left[ \frac{A_{-}^2-1}{v^{2/A_{-}}}%
\right] . 
\end{equation}
We note that $A_{-}$ is greater than one. Hence this expression diverges, in
a positive sense, in the approach to the Cauchy horizon, i.e. as $%
v\rightarrow 0$.

In the black hole case we use the ${\cal F}(v)$ from (\ref{ma2}) in the 
expression for the radiated power, to get 
\begin{equation}
\label{bpo}P(v)=\frac 1{48\pi M^2} 
\end{equation}
as expected.

In order to calculate the spectrum we recall that 
the number distribution of Minkowski particles observed on $\scrip$ is simply 
given
by the Bogoliubov coefficient
\begin{equation}
\label{bog}
\beta(\w',\w)~~ =~~ \int_{-\infty}^\infty {{d\hat U} \over {4\pi\sqrt{\w\w'}}}
e^{-i\w\hat U} e^{-i\w'{\cal G}(\hat U)} 
\end{equation}
(where the integral is performed over all of $\scrip$) as
\begin{equation}
\label{num}
_M \langle 0| N(\w)|0\rangle_M~~ =~~ \int_0^\infty d\w'
|\beta(\w',\w)|^2 \end{equation}
where $|0\rangle_M$ is the Minkowski vacuum.

Substituting for 
${\cal G}(\hat U)$ from (\ref{in2}) for the black hole case gives 
\begin{equation}
\label{bhb}
|\beta(\w',\w)|^2 = \frac{2M}{\pi \omega'}\frac{1}{e^{8\pi M\omega}-1}
\end{equation}
which is the expected Hawking spectrum. 

On the other hand, for the naked singularity case, we take the 
${\cal G}(\hat U)$
from (\ref{inv}) to get
\begin{equation}
\label{nak}
\beta(\w',\w)~~ =~~ {1 \over {2\pi}}\sqrt{\w\over \w'}  
\int_{-\infty}^{\hat U^o}
d\hat U e^{-i\w\hat U} e^{i\w' Q^{-A_{-}}(\hat U^o - \hat U)^{A_-}}.
\end{equation}

Changing variables to $z = (\hat U^o -\hat U)$, one has

\begin{equation}
\label{bet}
\beta(\w',\w)~~ =~~ {1 \over {2\pi}} \sqrt{\w \over \w'} e^{ - i\w \hat U^o}
\int_0^\infty dz e^{i\w z} e^{i \w' (z/Q)^{A_{-}}} 
\end{equation}

which gives

\begin{equation}
\label{be2}
|\beta(\w',\w)|^2~~ =~~ {1 \over {4\pi^2\w\w'}} |\sum_{k=0}^\infty
{{(iQ^{-A_{-}} \w' \w^{-A_{-}} e^{i\pi A_{-}/2})^k} \over {k!}}  
\Gamma (kA_{-}+1)|^2.
\end{equation}
This is a non-thermal spectrum, similar to the spectrum in the case of
self-similar dust collapse \cite{vaw}.

\section{Conclusion}

The divergence of the radiated power on the Cauchy horizon has been shown
by us \cite{suk} to be independent of the
assumption of self-similarity. The non-thermality of the spectrum is
due to the fact that not all of ${\cal I}^+$ is actually probed by 
infalling waves from ${\cal I}^-$.
We have, of course, implicitly assumed that all of ${\cal I}^+$
exists, i.e., that spacetime may be analytically continued beyond the Cauchy
horizon. This assumption is reasonable in that one does not expect a local 
collapse to terminate the universe. However, only a complete theory of 
quantum gravity can provide
the answer to the question of the existence of a complete ${\cal I}^+$,
as this depends on the final fate of the collapse. Therefore it is
interesting to consider
what might happen if the continuation beyond the Cauchy horizon is not
physically acceptable. If the Cauchy horizon is to be regarded as the
end point of spacetime, then $\hat{U}$ is not a good asymptotic coordinate.
However, one can transform to an asymptotically flat
retarded time and it can be shown \cite{vas} that the radiation
is thermal, but with a temperature that is different from the Hawking 
temperature. This is analogous to the marginally naked singularity treated by 
Hiscock et al. \cite{his}.

\end{document}